\documentclass[a4paper]{article}

\usepackage{INTERSPEECH2019}

\title{MOSNet: Deep Learning-based Objective Assessment for Voice Conversion}

\name{Chen-Chou Lo$^1$, Szu-Wei Fu$^2$, Wen-Chin Huang$^1$, Xin Wang$^3$, Junichi Yamagishi$^3$, \\ Yu Tsao$^2$, Hsin-Min Wang$^1$}

\address{
  $^1$Institute of Information Science, Academia Sinica, Taipei, Taiwan\\
  $^2$Research Center for Information Technology Innovation, Academia Sinica, Taipei, Taiwan\\
  $^3$National Institute of Informatics, Japan}
\email{\{lochenchou,unilight,whm\}@iis.sinica.edu.tw; \{jasonfu,yu.tsao\}@citi.sinica.edu.tw; \{wangxin,jyamagis\}@nii.ac.jp}

\begin{document}

\maketitle

\begin{abstract}

Existing objective evaluation metrics for voice conversion (VC) are not always correlated with human perception. Therefore, training VC models with such criteria may not effectively improve naturalness and similarity of converted speech. In this paper, we propose deep learning-based assessment models to predict human ratings of converted speech. We adopt the convolutional and recurrent neural network models to build a mean opinion score (MOS) predictor, termed as MOSNet. The proposed models are tested on  large-scale listening test results of the Voice Conversion Challenge (VCC) 2018. Experimental results show that the predicted scores of the proposed MOSNet are highly correlated with human MOS ratings at the system level while being fairly correlated with human MOS ratings at the utterance level. Meanwhile, we have modified MOSNet to predict the similarity scores, and the preliminary results show that the predicted scores are also fairly correlated with  human ratings. These results confirm that the proposed models could be used as a computational evaluator to measure the MOS of VC systems to reduce the need for expensive human rating.

\end{abstract}

\section{Introduction}
The quality quantification of generated speech has been a long-standing problem in speech synthesis, speech enhancement, and voice conversion (VC) systems. The evaluation of these systems reports both objective and subjective measurements. In the VC community, objective measures such as the Mel-cepstral distance (MCD) \cite{MCD} are widely used for automatically measuring the quality of converted speech. However, such metrics are not always correlated with  human perception as they measure mainly the distortion of acoustic features. Subjective measures such as the mean opinion score (MOS) and similarity score could represent the intrinsic naturalness and similarity of a VC system, but these type of evaluations are time-consuming and expensive as they need large number of participants to undergo listening tests and provide perceptual ratings.

Many assessment algorithms and models have been proposed to overcome the above-mentioned problem. For instance, in the speech enhancement field, the perceptual evaluation of speech quality (PESQ) \cite{PESQ} released by ITU-T is an intrusive assessment for measuring the quality of enhanced speech because a golden reference is needed for evaluation. There are several non-intrusive assessment metrics \cite{Cernak2005AnEO,PESQpredict, Remes2013ObjectiveEM, yoshimura,Fu, automos, nisq, PESQ2} for evaluating the quality of enhanced speech and synthetic speech. For instance, Fu \textit{et al.}\cite{Fu} proposed Quality-Net as a quality assessment model based on bidirectional long short-term memory (BLSTM) to predict the utterance-level quality of enhanced speech in a frame-wise manner. The high correlation between the predicted scores and the PESQ scores confirmed its effectiveness as a non-intrusive assessment model for speech enhancement. Yoshimura \textit{et al.}\cite{yoshimura} proposed a naturalness predictor for synthetic speech based on a fully connected neural network and a convolutional neural network (CNN) to predict the utterance-level and system-level MOS. The model was conducted on handcrafted features and trained with large-scale listening test ratings. It is worth noting that both \cite{yoshimura} and \cite{automos} have reported a high variance in the utterance-level human ratings since the listening test is subjective and listeners may provide variant ratings to the same utterance. Therefore, it is difficult for an assessment model to get a high correlation with human utterance-level ratings. Nevertheless, the system-level prediction would be relatively reliable. Previous works have shown the capability of neural networks in modeling human perception for enhanced synthetic speech. However, there is no such assessment model for VC systems. Our goal is to develop a speech naturalness and similarity assessment model for VC systems using large-scale listening test results of the Voice Conversion Challenge (VCC) 2018 \cite{VCC2018}.



In this paper, we present a novel deep learning based end-to-end speech naturalness assessment model, termed MOSNet. To develop such objective measures to model and align with human subjective ratings, we investigated the convolutional neural network (CNN), bidirectional long short-term memory (BLSTM), and CNN-BLSTM, as these architectures have shown their capability to model human perception. We used such architectures for extracting valuable features and then used fully connected (FC) layers to predict the corresponding naturalness scores. In virtue of the capacity of neural networks and the large-scale human naturalness evaluations of the VCC 2018, the MOS prediction of our naturalness assessment model achieved high correlation with human MOS ratings at the system level and fair correlation at the utterance level. In addition, we modified MOSNet to predict the similarity scores, and the preliminary results  showed that the predicted similarity scores are fairly correlated with human similarity ratings. The experimental results showed that our proposed models have high capability to measure speech naturalness and similarity of VC systems. As per our knowledge, this is the first deep learning-based speech quality and similarity assessment model for VC.

The paper is organized as follows: The data and its distribution from VCC 2018 are described in Section 2. The proposed models are presented in Section 3. The experiments and results are discussed in Section 4. Finally, the conclusions and future work are presented in Section 5.

\section{The Voice Conversion Challenge Evaluation Data}

\subsection{The Voice Conversion Challenge 2018}

The Voice Conversion Challenge (VCC) 2018, the second edition of VCC, is a large-scale voice conversion challenge. The VCC 2018 corpus was prepared by selecting a part of speakers from the device and production speech (DAPS) dataset \cite{DAPS} recorded by professional US English speakers in a clean and noise-free environment. Participants in the challenge needed to use their VC systems trained by parallel or non-parallel training data to convert speech signals from a source speaker to a target speaker. All the parallel and non-parallel VC systems participating in the challenge had been evaluated in terms of naturalness and similarity scores via crowd-sourcing listening tests.

The evaluation of the VCC 2018 was as follows: There were 2,572 evaluation sets, each consisting of 44 utterances. A total of 113,168 human evaluations fully covered 28,292 submitted audio samples. Each audio sample was rated by 4 listeners. 113,168 evaluations were split into 82,304 naturalness assessments and 30,864 speaker similarity assessments. 82,304 naturalness evaluations covered 20,580 submitted utterances with MOS ranging from 1 to 5, with lowest score of 1 and highest score of 5. The detailed specification of the corpus, listeners and evaluation methods can be found in \cite{VCC2018}. We took the average score of the four MOS ratings for each utterance as its ground-truth score.

\subsection{Data along with its distribution and predictability}

\begin{figure}[t]
  \centering
  \includegraphics[width=\linewidth]{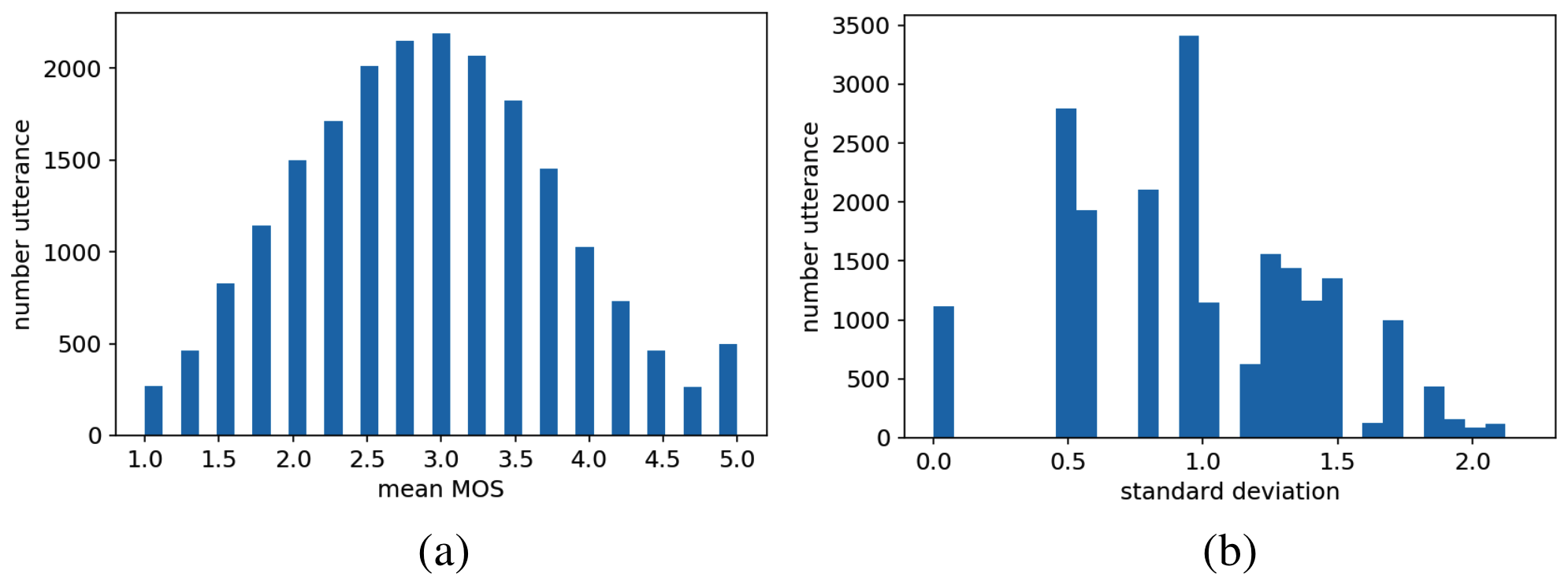}
  \vspace{-5mm}
  \caption{Histograms of the mean (a) and standard deviation (b) of four MOS ratings for each utterance in the VCC 2018.}
  \label{fig:data_distribution}
\end{figure}

The histograms of the mean and standard deviation of the four MOS ratings for each utterance are shown in Figure~\ref{fig:data_distribution}. It can be seen that the distribution of the mean MOS is closer to a Gaussian, and the mean MOS values are concentrated around 3.0. However, for about half of the submitted utterances, the standard deviation of the four MOS ratings is greater than 1, suggesting a higher degree of variation in the scores. This is expected because the perceptual ratings for the same utterance depend on the listeners' personal experiences and preferences when conducting listening tests. Therefore, we established that it is important to consider the intrinsic predictability of data and the inherent correlation among listeners. In this study, we used the bootstrap method in \cite{yoshimura, bootstrap, bootstrap2} to verify the inherent predictability of human evaluations in the VCC 2018.

We took 1,000 replications to estimate the MOS correlation between each subset and the whole dataset. Note that the MOS evaluations of natural speech was excluded. For each replication, we randomly sampled 134 listeners from a total of 267 listeners to measure their mean MOS as MOS$_{sub}$. Then, MOS$_{sub}$ was compared to MOS$_{all}$ (computed using the entire set of MOS ratings) in terms of the linear correlation coefficient (LCC) \cite{LCC}, Spearman's rank correlation coefficient (SRCC) \cite{SRCC} and mean square error (MSE). The average LCC, SRCC, and MSE values are shown in Table~\ref{tab:inherent_correlation}. Hence, it became clear that LCC and SRCC are quite high at the system level but lower at the utterance level. MSE shows a similar trend. The results indicate that although subjective perceptual ratings of different listeners vary at the utterance level, they have good consistency at the system level. This analysis shows that although the MOS at the system level is predictable, but, the MOS at the utterance level can be predicted only up to a certain extent, although not as good as the prediction at the system level.

\begin{table}
  \caption{Inherent correlation of human evaluations in the VCC 2018 at the utterance and system levels.}
  \label{tab:inherent_correlation}
\centering
\footnotesize
\tabcolsep 4.5pt
\vspace{-2mm}
\begin{tabular}{cccc} 
\toprule
\textbf{Level} & \textbf{LCC} & \textbf{SRCC} & \textbf{MSE}  \\ 
\hline
Utterance      & 0.805        & 0.806         & 0.396         \\
System         & 0.994        & 0.978         & 0.005         \\
\bottomrule
\end{tabular}
\end{table}

\section{MOSNet}

This paper proposes a deep learning-based objective assessment to model human perception in terms of MOS, referred to as MOSNet. Raw magnitude spectrogram is used as the input feature and three neural network-based models, namely CNN, BLSTM and CNN-BLSTM are used to extract valuable features from the input features of the fully connected (FC) layers and pooling mechanism to generate predicted MOS. In the following sections, we will detail each component of MOSNet.

\subsection{Model details}

The detailed configuration of different architectures of MOSNet is shown in Table~\ref{tab:model_config}, including CNN, BLSTM, and CNN-BLSTM. The BLSTM architecture is the same as the one used in Quality-Net \cite{Fu}. In virtue of forward and backward processing through time, BLSTM has the ability to integrate long-term time dependencies and sequential characteristics into representative features. CNN has been widely used to model time series data and yielded satisfactory performance. CNN expands its receptive field through stacking more convolutional layers. The architecture of CNN used in this study has 12 convolutional layers,  its receptive field for each neuron in the last convolutional layer is 25 frames (around 400 ms in the time-scale). We believe that, by considering the segment of 25 frames, the MOSNet can capture more temporal information to predict the quality scores. Recent studies have confirmed the effectiveness of combining CNN and RNN (BLSTM) for enhancement \cite{CLDNN, tan2018convolutional}, classification \cite{CRDNN, lu2018temporal}, and recognition \cite{speechCLDNN, wang2018gated} tasks. Thus, we also investigate the CNN + BLSTM architecture for feature extraction in the MOSNet, which is referred to as CNN-BLSTM in Table~\ref{tab:model_config} and in subsequent discussions. With the extracted features, we use two FC layers to regress frame-wise features into a frame-level scalar to indicate the naturalness score of each frame. Finally, a global averaging operation is applied to the frame-level scores to obtain the utterance-level naturalness score.

\begin{table}
\centering
\caption{Configuration of different model architectures. The convolutional layer parameters are denoted as “conv\{receptive field size\}-\{number of channels\}/\{stride\}.” The ReLU \cite{relu} activation function after each convolutional layer is not shown for brevity. N is for the number of frames.}

\label{tab:model_config}
\footnotesize
\tabcolsep 4.5pt
\vspace{-2mm}
\begin{tabular}{c|c|c|c} 
\toprule
model                                                     & \textbf{BLSTM }                                                  & \textbf{CNN }                                                                    & \textbf{CNN-BLSTM }                                                                                                               \\ 
\hline
\begin{tabular}[c]{@{}c@{}}input\\layer \end{tabular}     & \multicolumn{3}{c}{input (\textbf{\textit{N X 257 mag spectrogram}}) }                                                                                                                                                                                                                  \\ 
\hline
\begin{tabular}[c]{@{}c@{}}conv.\\layer \end{tabular}     & \multicolumn{1}{l|}{}                                            & \multicolumn{2}{c}{\begin{tabular}[c]{@{}c@{}}\\$ \left \{\begin{aligned}conv3-(channels)/1 \\conv3-(channels)/1 \\conv3-(channels)/3\end{aligned}\right \} X 4$\\ \\ $channels$ = [16, 32, 64, 128] \end{tabular}}  \\ 
\hline
\begin{tabular}[c]{@{}c@{}}recurrent\\layer \end{tabular} & BLSTM-128                                                        &                                                                                  & BLSTM-128                                                                                                                         \\ 
\hline
\begin{tabular}[c]{@{}c@{}}\\FC \\layer\end{tabular}      & \begin{tabular}[c]{@{}c@{}}FC-64, \\ReLU,\\dropout \end{tabular} & \begin{tabular}[c]{@{}c@{}}FC-64,\\ReLU,\\dropout \end{tabular} & \begin{tabular}[c]{@{}c@{}}FC-128,\\ReLU,\\dropout \end{tabular}                                                 \\ 
\cline{2-4}
                                                          & \multicolumn{3}{c}{FC-1 (\textbf{\textit{frame-wise scores) }} }                                                                                                                                                                                                                        \\ 
\hline
\begin{tabular}[c]{@{}c@{}}output\\layer \end{tabular}    & \multicolumn{3}{c}{average pool (\textbf{\textit{utterance score}}) }                                                                                                                                                                                                                   \\
\bottomrule
\end{tabular}
\vspace{-2mm}
\end{table}



\subsection{Objective function}

As described in the previous section, we formulate MOS prediction as a regression task. The input of the MOSNet is a sequence of spectral features, extracted from a speech utterance. The MOS evaluations of VCC 2018 are used as the ground-truth to train the model. Fu \textit{et al.}\cite{Fu} stated that by using frame-level prediction errors in the objective function, utterance-level predictions will be more correlated with human ratings. Thus, we formulate the objective function for training the MOSNet as:

\begin{equation}
    O = \frac{1}{S}\sum_{s=1}^{S}[(\hat{Q_{s}}-Q_{s} )^{2} +\frac{\alpha}{T_{s}}\sum_{t=1}^{T_{s}}(\hat{Q_{s}}-q_{s,t} )^{2} ]
      \label{obj_funcion}
\end{equation}

\noindent where $\hat{Q_{s}}$ and $Q_{s}$ denote the ground-truth MOS and predicted MOS for the $s$-th utterance, respectively, $\alpha$ is the weighting factor, $q_{s,t}$ denotes the frame-level prediction at time $t$, $T_{s}$ is the total number of frames in the $s$-th utterance, and $S$ denotes the number of training utterances. Notably, the objective function in Eq. (\ref{obj_funcion}) combines the utterance-level MSE and frame-level MSE. In \cite{Fu}, the weighting factor ($\alpha$) was used to mitigate severe MSE variations across frames in a speech enhancement task. Our preliminary experiments show that the quality of converted speech is much more stable across frames compared to the quality of enhanced speech. Specifically, the converted speech generated from a high MOS VC system typically produces high frame-wise MOSs, and vice versa. Thus, we set the weighting factor $\alpha$ to 1 in this study. To compute the frame-level MSE, the ground-truth MOS is used for all the frames in the speech utterance. From the experiments, it can be established that the frame-level MSE helps MOSNet converge with better prediction accuracy, which will be discussed in the next section. 

\begin{table}
\centering
\caption{Utterance-level and system-level prediction results for different models, where the subscript denotes the batch size.}
\label{table:model_result}
\footnotesize
\tabcolsep 2pt
\vspace{-2mm}
\begin{tabular}{lllllll} 
\toprule
                                   & \multicolumn{3}{c}{\textbf{\textit{utterance-level}}}                                                   & \multicolumn{3}{c}{\textbf{\textit{system-level}}}                                                       \\ 
\hline
\multicolumn{1}{c}{\textbf{$\textbf{Model}_{\textit{batchsize}}$}} & \multicolumn{1}{c}{\textbf{LCC}} & \multicolumn{1}{c}{\textbf{SRCC}} & \multicolumn{1}{c}{\textbf{MSE}} & \multicolumn{1}{c}{\textbf{LCC}} & \multicolumn{1}{c}{\textbf{SRCC}} & \multicolumn{1}{c}{\textbf{MSE}}  \\ 
\hline
BLSTM$_{\textit{1}}$ \cite{Fu}                            & 0.511                           & 0.484                           & 0.604                           & 0.826                           & 0.808                            & 0.165                            \\
BLSTM$_{\textit{16}}$                             & 0.487                           & 0.453                            & 0.658                           & 0.818                           & 0.797                            & 0.190                            \\
BLSTM$_{\textit{64}}$                               & 0.251                           & 0.254                            & 0.803                           & 0.412                           & 0.427                            & 0.404                            \\ 
\hline
CNN$_{\textit{1}}$                             & 0.638                           & 0.587                            & \textbf{0.486}                  & 0.945                           & 0.875                            & 0.058                            \\
CNN$_{\textit{16}}$                                & 0.620                           & 0.573                            & 0.512                           & 0.944                           & 0.890                            & 0.067                            \\
CNN$_{\textit{64}}$                                & 0.624                           & 0.585                            & 0.522                           & 0.946                           & 0.872                            & 0.057                            \\ 
\hline
CNN-BLSTM$_{\textit{1}}$                          & 0.584                           & 0.551                            & 0.634                           & 0.951                           & 0.873                            & 0.135                            \\
CNN-BLSTM$_{\textit{16}}$                             & 0.607                           & 0.569                            & 0.540                           & 0.944                           & \textbf{0.897}                   & \textbf{0.055}                   \\
CNN-BLSTM$_{\textit{64}}$                              & \textbf{0.642}                  & \textbf{0.589}                   & 0.538                           & \textbf{0.957}                  & 0.888                           & 0.084                           \\
\bottomrule
\end{tabular}
\end{table}

\section{Experiments}

In this section, different model architectures will be evaluated and the results at the utterance and system levels will be discussed. We explain our implementation details first. The entire set of 20,580 speech samples (along with the corresponding MOSs) were divided into 13,580, 3,000 and 4,000 samples for training, validation and testing, respectively. All speech samples were down-sized to 16 kHz. For feature extraction, we conducted short-time Fourier transform (STFT) of 512 sample points (i.e., 32 ms frame size) every 256 sample points (i.e., 16 ms frame shift), which resulted in a sequence of frame-based spectral feature of 257 dimensions for each utterance. For the MOSNet, the dropout \cite{dropout} rate was set to 0.3. The model was trained by the Adam \cite{adam} optimizer with a learning rate of 0.0001, without batch normalization. We applied early stopping based on the MSE of the validation set with 5 epochs patience.


\subsection{Comparison of different model architectures}

First, we intend to compare the prediction performance of different models. Table~\ref{table:model_result} shows the LCC, SRCC and MSE values of these models with different batch sizes at the utterance and system levels. The model BLSTM$_{\textit{1}}$ used in \cite{Fu} is considered to be the baseline system. From Table~\ref{table:model_result}, BLSTM$_{\textit{1}}$ yielded 0.511 in terms of LCC, and its performance dropped drastically when the batch size increased, as zero-padding might affect the prediction performance. Next, the CNN-based models yielded more stable results when the batch size increased and boosted the LCC from 0.511 to 0.638. Among all the models, CNN-BLSTM$_{\textit{64}}$ achieved the best LCC of 0.624, indicating that the combination of CNN (for feature extraction) and BLSTM (considering time dependencies) could effectively extract features from the spectrogram to perform MOS prediction. CNN-BLSTM$_{\textit{64}}$ also achieved the highest LCC of 0.957 at the system level, which is very close to the LCC of 0.994 for human evaluation shown in Table~\ref{tab:inherent_correlation}.

\subsubsection{Evaluated results at the utterance level}

\begin{figure}[t]
  \centering
  \includegraphics[width=\linewidth]{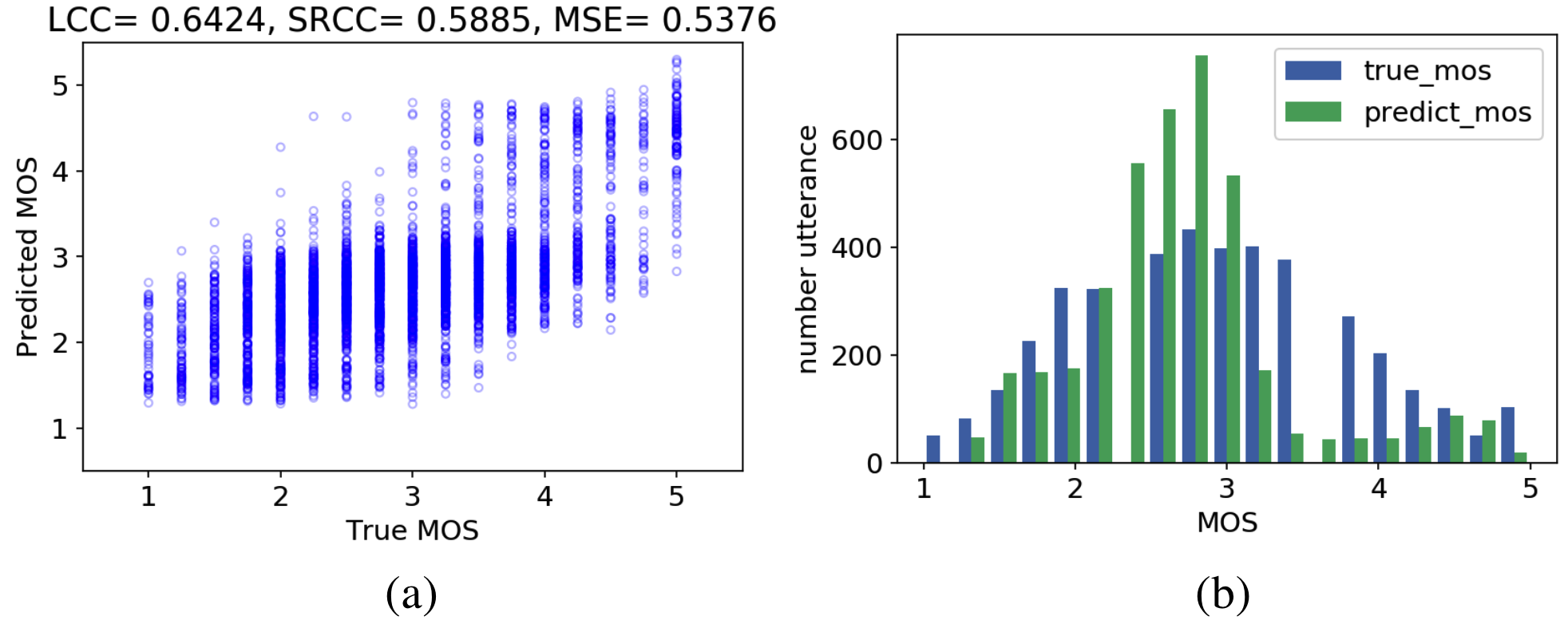}
  \vspace{-3mm}
  \caption{Scatter plot (a) and histogram (b) of the utterance-level predictions of CNN-BLSTM$_{\textit{64}}$.}
  \label{fig:utterance_mos}
\end{figure}

Figure~\ref{fig:utterance_mos} shows a scatter plot and histogram of the utterance-level predictions of CNN-BLSTM$_{\textit{64}}$. As seen from Figure~\ref{fig:utterance_mos} (a), the model avoided low predictions although there were quite a few training samples in the low MOS, and the predictions were mainly distributed around scores of 2 to 3. Figure~\ref{fig:utterance_mos} (b) shows that MOSNet rarely predicted MOS between 3.5 to 4 and mostly predicted MOS between 1.5 and 3.5. As the average MOS of the submitted systems fell mainly between 2 to 3.5 at the system level, it is reasonable for the model to avoid prediction of  such low and high scores. This is a common limitation when using MSE-based objective functions for the data is in a Gaussian distribution. It can be improved by using other objective functions, which can be a scope of future research.

\begin{figure}[t]
  \centering
  \includegraphics[width=\linewidth]{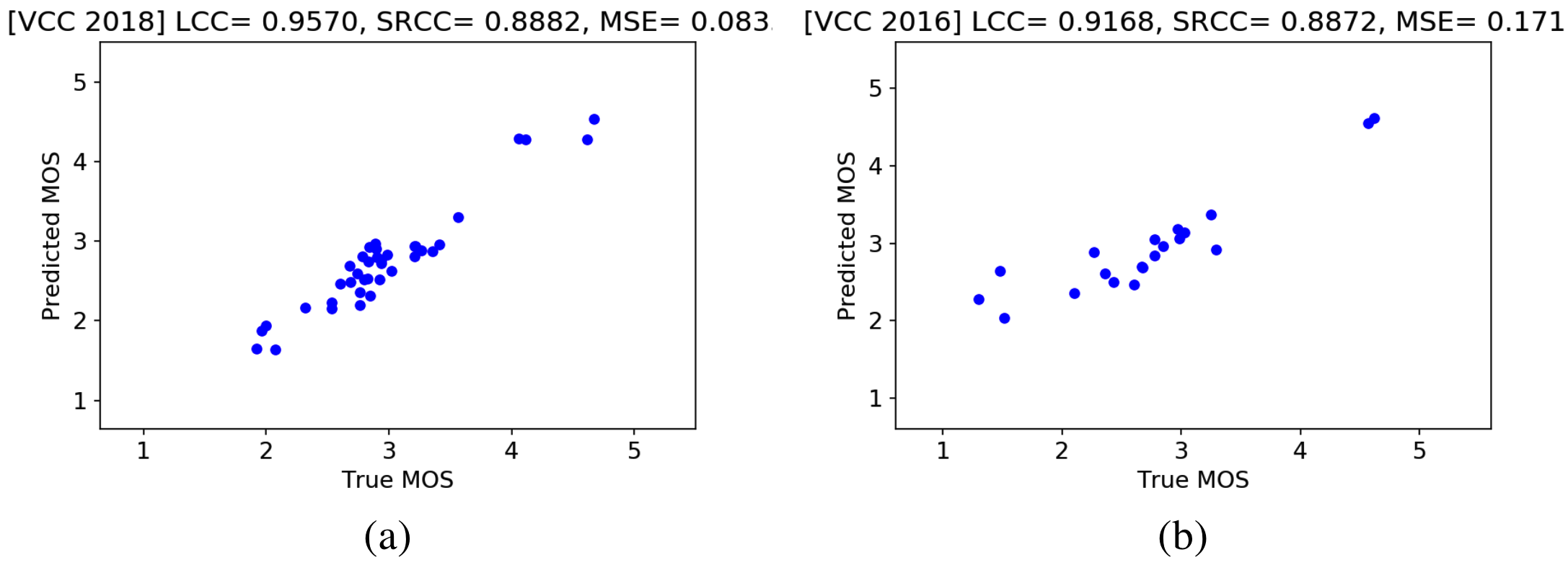}
  \vspace{-3mm}
  \caption{Scatter plot of system-level predictions on the testing set in (a) the VCC 2018 and (b) the VCC 2016.}
  \label{fig:system_mos}
\end{figure}

\subsubsection{Evaluated results at the system level}

Figure~\ref{fig:system_mos} (a) shows the scatter plot of system-level predictions by CNN-BLSTM$_{\textit{64}}$ for the VCC 2018. System-level predictions were highly correlated with human evaluations. The results shown in Figure~\ref{fig:system_mos} (a) and Table~\ref{table:model_result} confirmed the effectiveness of the proposed MOSNet as a substitute for human listeners to evaluate VC systems.

To test the generalization of MOSNet, we applied CNN-BLSTM$_{\textit{64}}$ trained by the VCC 2018 data to the VCC 2016 data. The scatter plot of system-level predictions is shown in Figure~\ref{fig:system_mos} (b). Although, the audio samples in both VCC 2016 and VCC 2018 came from the DAPS dataset, the selected speakers for testing of  VCC 2016 were different from those of VCC 2018. The participating VC systems for the two challenges were also different. As the pair correspondence results for audio samples were not accessible, we could only report the system-level prediction results. Figure~\ref{fig:system_mos} (b) shows that MOSNet achieved a sufficiently high correlation of 0.917 for system-level predictions, thereby confirming its generalization ability in a training-testing mismatched scenario.

\subsection{The effect of the frame-level MSE}

\begin{figure}[t]
  \centering
  \includegraphics[width=0.875\linewidth]{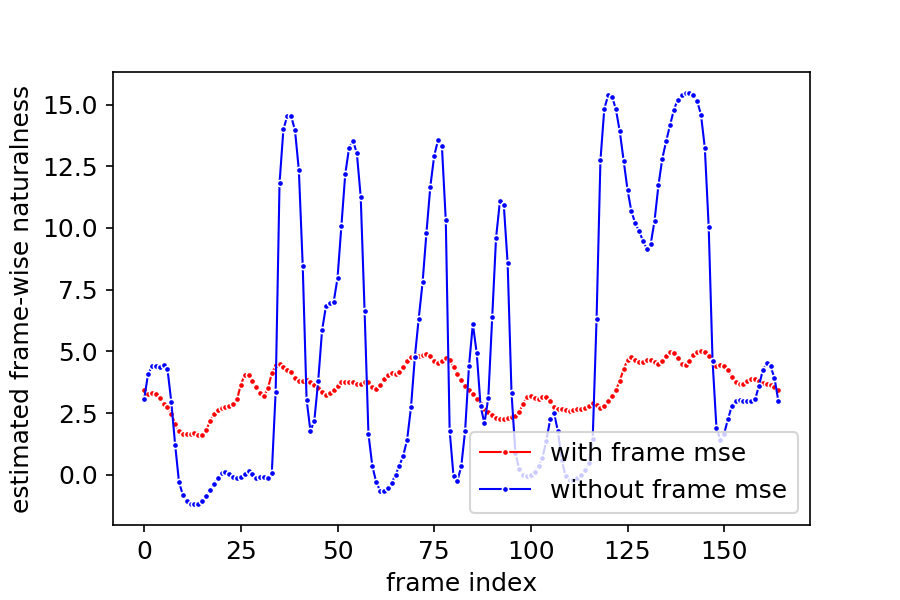}
  \vspace{-3mm}
  \caption{Frame-wise MOS predictions of CNN-BLSTM$_{\textit{64}}$.}
  \label{fig:frame_score}
\end{figure}

Here, we intend to investigate the effect of the frame-level MSE used in the objective function in Eq. (\ref{obj_funcion}). Figure~\ref{fig:frame_score} and Table~\ref{tab:frame_result} shows the effect of training with or without the frame-level MSE on the CNN-BLSTM$_{\textit{64}}$ model (the best system in Table~\ref{table:model_result}). As seen in Figure~\ref{fig:frame_score}, the estimated frame-wise scores vary greatly when the model was trained without the frame-level MSE, but are relatively stable when the model was trained with the frame-level MSE. Unstable frame-wise predictions affect the final utterance-level predictions. Table~\ref{tab:frame_result} shows that with frame-level MSE, the LCC of utterance-level predictions was 0.6424, and it dropped significantly to 0.5604 when the model was trained without the frame-level MSE.

\begin{table}
  \caption{The effect of frame-level MSE on the utterance-level predictions of CNN-BLSTM$_{\textit{64}}$.}
  \label{tab:frame_result}
  \centering
  \footnotesize
  \tabcolsep 4.5pt
  \vspace{-2mm}
  \begin{tabular}{llll}
    \toprule
    \textbf{}      & \textbf{LCC}   & \textbf{SRCC}  & \textbf{MSE}               \\
    \midrule
      with frame MSE & \textbf{0.642} & \textbf{0.589} & \textbf{0.538}\\
      without frame MSE & 0.560 & 0.528 & 2.525\\
    \bottomrule
  \end{tabular}
\end{table}



\begin{table}
\centering
\caption{Similarity prediction results of the modified MOSNet model.}
\label{tab:sim_result}
\footnotesize
\tabcolsep 4.5pt
\vspace{-2mm}
\begin{tabular}{c|ccccc} 
\toprule
 \textbf{model}  & \textbf{Level}  & \textbf{ACC}    & \textbf{LCC}    & \textbf{SRCC}   & \textbf{MSE}     \\ 
\hline
CNN              & utterance       & 0.696           & \textbf{0.453}  & \textbf{0.455}  & 0.197            \\
(scalar)         & system          & \textbf{0.701}  & 0.394           & 0.395           & \textbf{0.195}   \\ 
\hline
CNN              & utterance       & 0.670           & 0.329           & 0.329           & 0.336            \\
(2 classes)      & system          & 0.674           & 0.292           & 0.292           & 0.326            \\
\bottomrule
\end{tabular}
\end{table}

\subsection{Experiments on similarity prediction}

The similarity in the measurement of the converted speech and the target speech is another key indicator of the success of VC systems. In addition to predicting the MOS, we also tried to extend the MOSNet to predict the similarity scores. As there were two input audio samples at the same time, we applied a shared CNN (with the same architecture as described in Table~\ref{tab:model_config}) to extract features into latent vectors for both the input samples. The paired vectors were combined to form a concatenated feature vector and inferred by two FC layers and a SoftMax layer to generate the final predicted score. Two models with different types of output were created: the first model had a scalar output, and the output of the second model was a two-class vector.

In VCC 2018, there were 30,864 evaluated similarity scores. Each similarity score was obtained by evaluating a pair of speech samples with 4 levels. We merged scores 1 and 2 as label 1 (same speaker), and scores 3 and 4 as label 0 (different speaker). The dataset was divided into 80\% and 20\% for training and testing, respectively. The similarity prediction results of the two models are shown in Table~\ref{tab:sim_result}. From the table, we found that when the model is trained with one scalar output, it can achieve an accuracy of 69.6\%, which is notably higher than 66.9\% yielded by the two-class model. The results in Table~\ref{tab:sim_result} confirmed that the modified MOSNet can be used to predict similarity scores with fair correlation values to human ratings. 

\section{Conclusions}

This paper presented a deep learning-based quality assessment model for the VC task, referred to as MOSNet. Based on large-scale human perceptual MOS evaluation results from VCC 2018, our experimental results show that MOSNet yields predictions with a high correlation to human ratings at the system level and a fair correlation at the utterance level. We have shown decent generalization capability of MOSNet by applying the model trained with the VCC 2018 data to the VCC 2016 data. Moreover, with a slight modification, MOSNet can fairly predict the similarity scores of the converted speech relative to the target speech. As per our knowledge, the proposed MOSNet is the first end-to-end speech objective assessment model for VC. In future, we will consider the human perception theory and improve the model architecture and objective function of MOSNet to attain improved correlation with human ratings. \\
\noindent\textbf{Acknowledgements}: We are grateful to the VCC for releasing the evaluation data of the VCC 2018.

\bibliographystyle{IEEEtran}

\bibliography{mosnet}

\end{document}